# A Novel Approach for Canvas Accessibility Problem in HTML5

**Paniz Alipour Aghdam[1], Reza Ravanmehr[2]**

[1] Department of Computer Engineering, Malayer Branch, Islamic Azad University
Malayer, Iran
alipourpaniz@gmail.com

[2] Department of Computer Engineering, Central Tehran Branch, Islamic Azad University
Tehran, Iran
r.ravanmehr@iauctb.ac.ir

### Abstract

Canvas is a pixel-based inherently inaccessible element in HTML5. Therefore web users with vision disabilities cannot benefit from Canvas and its desired semantics and functionality. Regarding to the Canvas application in designing interactive graphical user interface, vision-impaired users may miss important information on web sites.

This paper utilizes the content-based image retrieval (CBIR) technique as well as code mapping embedded in a Firefox extension to present a novel approach in order to make Canvas interactive user interface accessible. This extension replaces Canvas with an accessible equivalent HTML environment. Unlike previously done works on Canvas accessibility, the proposed approach does not impose any rules on developers and designers during Canvas design.

## 1. Introduction

Nowadays, the impact of web environment in our lives is growing gradually more. Due to increase in the usage of visual contents (e.g. graphics, pictures), or items, which are based on visual perception (e.g. tables, diagrams, etc.), as well as advent of technologies with accessibility challenges in web environment (e.g. Canvas), visually impaired people encounter some accessibility problems. Currently, visually impaired people access the web sites by using screen readers that is a software tool capable of reading the web page aloud (with a synthesized voice); Interaction is possible using Braille keyboards. This group of users won't be able to access the mentioned contents and technologies if their alternatives are not provided, or rules of accessibility are not observed [1-4].

Canvas is one of the newly introduced elements in HTML5 which as mentioned before is one of the challenges in web accessibility. This element is pixel-based, consequently inherently inaccessible especially for visually-impaired people. That is due to inability of assistive technologies in accessing the semantics of what is drawn on Canvas.

Even though all users cannot make full use of Canvas elements, it does not mean that such features should be omitted entirely. Instead, alternate mechanisms need to be provided. This element enables dynamic rendering of graphics such as graphs, bitmap images, animations and games using scripts. Canvas can be used for UI construction for web application too [6-9].

For Images and animated graphic demos it is believed that a text could suffice. However, for truly interactive application-like content using an approach for making Canvas accessible is essential. There are no methods to assign roles, states and properties to areas of a canvas that represent widgets, controls or links. Therefore, the developer who wants to create an accessible interface has to duplicate the content of the canvas interface using HTML elements, which provide native roles, states and properties [7, 9].

When authors use the canvas element, they must also provide content, which conveys essentially the same function or purpose as the bitmap canvas when presented to user. This content may be placed as content of the canvas element. The contents of the canvas element, if any, are the element's fallback content [7, 8].

In general., "fallback content" (i.e., content included to benefit users whose web browsers don't support a particular feature or format) is not appropriate as accessibility solution since users with disabilities are more likely to be using as modern and powerful browsers as anybody else [7,8]. The case with canvas is similar; while some sort of DOM-based alternative content is a possible solution to canvas accessibility concerns, it seems that won't work without some modifications to the spec and/or canvas API. In order to make Canvas accessible many researchers are working on a solution. Researchers proposed two main approaches for this purpose -

Navigable DOM and the other one which is a version of a client-side image map.

In image map solution, each actually created HTML element in the DOM could be attached to a certain pixel region of a Canvas element. By Image map solution, Canvas was not only keyboard focusable, but also accessible by assistive technology. However this solution had some limitations since it imposed additional requirements on implementers. Moreover, it was confined to some basic image map shapes [10, 11].

The Navigable DOM behavior causes the browser exposes the content inside a Canvas element to assistive technology users even when the browser supports Canvas [8]. Thus, Canvas can be made keyboard-focusable. Moreover, Aria roles can be helpful in exposing Canvas contents to assistive technologies. Although, this behavior has not been completely implemented on all browsers that support Canvas, the recently released Firefox 13 implements the HTML5 specification's Canvas element fallback concept. So that the content inside a Canvas elements start and end tags is exposed to assistive technology users even when the browser supports Canvas. Considering this new progression in Canvas accessibility, again it imposes rules and limitations on designers and developers which overlaping them can causes inaccessibility of Canvas. That is the problem which in our new approach was attempted to be resolved [8, 12].

Both aforementioned studies and other ones which aim to make interactive canvas accessible impose some rules and limitations on authors and web developers. Consequently, ignoring these rules can cause inaccessibility of Canvas. The main objective of this paper is to present a novel solution to make interactive Canvas accessible.

The proposed solution does not impose any rules or limitations on web developers or authors, and this is the advantage of the solution in comparison with other ones.

In this approach, an extension has been implemented in Firefox browser. The extension converts user interface designed on Canvas to accessible equivalent HTML User interface in web environment.

Moreover, extensions can be implemented on all browsers. Furthermore, there is no need for user interference. Therefore it can be taken into account as another advantage of the proposed solution.

The rest of the paper is organized as follows: brief reviews of the proposals, which aimed to make Canvas accessible, investigate the proposed solution, examining the results and finally, the conclusion.

## 2. Related Work

The start point of Canvas accessibility researches was Bespin project in 2009. Bespin was an open extensible web-based framework for code editing [9, 13]. Moreover, it was one of the main proposals aimed to add a use map attribute to Canvas.

This approach involves allowing the use map attribute on Canvas. It allows any element in the DOM to serve as an image map area. In this way, an actual button element can be created in the DOM and attached to a certain pixel region of a Canvas element, in which case the latter region would be keyboard focusable and announced to assistive technology. While adding Use map, attribute had some advantages such as simply adding interactive area on a Canvas element as well as making Canvas keyboard focusable and accessible to assistive technology. Also, it had some limitations such as imposing additional requirements on designers, and also restriction to basic image map shapes. These limitations caused this proposal to be rejected [14, 15].

The Navigable DOM concept provides a canvas alternative using a tree of equivalent elements as descendants of the canvas content. When authors use the Canvas element, they must also provide content that, when presented to the user, conveys essentially the same function or purpose as the bitmap Canvas. Moreover, when a Canvas element represents embedded content, the user can still focus on descendants of the Canvas element (in the fallback content). Therefore, authors can make an interactive Canvas keyboard-focusable [8, 9].

Also in Navigable DOM, getting help from Aria roles can expose elements to assistive technologies. However, this behavior has not been completely implemented on all browsers that support Canvas [8].

In addition to these methods, other proposals have been presented, which are based on Navigable DOM proposal acceptance.

Nonav and adom are two proposal instances, both of which present Boolean attribute for different purposes on Canvas. Nonav Boolean attribute has a potential valid use, hiding DOM children of the Canvas from assistive technology and keyboard navigation when this content is intended solely as fallback for browsers that do not support Canvas at all [16].

Some of its positive points are:

- Providing a simple mechanism for authors to use to ensure that content in the canvas sub-tree is not available to users of user agents that support Canvas, when the content has been authored for the situation when Canvas is not supported.

- Stating clearly what authors are required to do to provide accessible content when the canvas sub-tree is used as a container for content providing interaction and semantics in browsers that support Canvas.

- Stating clearly how user agents, that support accessibility API, should process the sub-tree to support the accessibility API.

- Clarifying how focus management should be supported in the canvas sub-tree.

- Disambiguating the case where canvas sub-tree content is not fallback [15].

Because of the existing alternate approaches for nonav, HTML Working Group rejected this proposal. The alternate approach is aria-hidden attribute for Canvas. So that by setting this attribute with true, all the sub-nodes of Canvas would be invisible.

Adom was a sign for indicating the type of Canvas sub-tree, whether it is fallback sub-tree or accessible sub-tree. The main challenge is that the author also uses the same <canvas> sub-tree to represent fallback content. The fallback sub tree differs from the accessible <canvas> sub-tree in that the accessible <canvas> sub-tree is designed to have its rendering directly drawn to the <canvas> whereas the fallback tree is not.

Due to mapping Canvas fallback content to accessibility API services by default, this proposal was rejected as well [15].

Overall, these proposals impose some rules and limitations on web designers and developers to make Canvas accessible. Therefore overlooking the rules causes inaccessibility of Canvas element for vision-impaired people.

# 3. Browser Extension a Solution for Accessibility

The proposed approach aims at converting interactive Canvas content into an accessible environment in web sites.

The main idea behind the approach is designing and implementing an extension in Firefox browser. So that changes in DOM immediately can be presented on the browser using browser layout engine.

The Concluded results of statistics and survey of two valid websites, Statcounter and Webaim, are the main reasons to choose Firefox for implementing the extension.

Webaim website survey reasoned that vision-impaired people mostly use Internet explorer and Firefox. Moreover, Canvas has been just supported by the IE, 9th version. Hence Firefox looks a better platform for implementing the idea [16, 17].

The implemented extension on Firefox uses image processing science to access the pixel-based Canvas.

In fact, the extension utilizing content-based image retrieval approach, which is one of the image recognition techniques, recognizes designed elements on Canvas [18]. Detecting different kinds of user interface elements on Canvas is performed by Content-based image retrieval technique. Thus, considering and putting correspondent objects in extension database seems appropriate. The other aspect of the extension is converting and mapping Canvas methods and properties to equivalent and common JavaScript methods and properties.

Keyboard is the most-used tool by blind people to interact with computer; therefore the extension adds functionality to the substitute environment of Canvas. Keyboard accessibility in superseded environment of Canvas is another advantage of the proposed solution.

Overall, this extension aims at converting the graphical application-like content Canvas to a HTML environment, which represents the same Canvas functionalities in accessible manner. Moreover, it adds useful attribute of WAI-Aria as well as keyboard accessibility to the superseded environment to improve the accessibility for blind people who use assistive technologies.

## 3.1 The Extension System

The operations of extension will be started when the webpage and its Canvas loaded completely. The Canvas element is located by document object and then the extension object detection stage is started. Canvas has a method which allows modifying Canvas pixels. Furthermore, by using this method, accessing different information of each pixel such as color and alpha value is possible. This method is *getImageData* () which has four arguments:

*Context. GetImageData(x, y, width, height)*

*(x , y)*: Relative coordinates to origin of the pixel region coordinates.

Width: Width of the pixel region.

Height: Height of the pixel region.

This method returns 2-D context rendering object. This object has three properties such as width and height of the pixel region of accessing region and data, which is a Canvas pixel array. This array contains information about all accessing pixel regions of Canvas [8].

In next step, pre-processing operations are done for image recognition, which includes noise reduction and edge detection. Zero-crossing method is used for edge detection, which its basis is Laplacian of Gaussian filter. For estimation of Laplacian on discrete image, it is required to use discrete convolution kernel.

The kernel used is (Table 1):

Table 1. Kernel Used for convolution

| 1 | 1 | 1 |
|---|---|---|
| 1 | -8 | 1 |
| 1 | 1 | 1 |

Next step is region labeling or region coloring by FloodFilling algorithm. This algorithm pieced together neighboring pixels in a stepwise manner to build regions in which all pixels within that region are assigned a unique number (label) for identification. FloodFill has three approaches: 1.recursive version 2.iterative depth-first version 3.iterative breadth-first version.

Since the depth of recursion is proportional to the size of the region, stack memory is quickly exhausted. Therefore, this method is risky and only practical for very small images. In depth-first approach, pixel tree is traversed by use of its own dedicated stack. In breadth-first approach, the data structure used to hold the unvisited pixels coordinates is, queue instead of stack. In comparison of these three methods, the memory requirements of breadth-first version is much lower than the depth-first version [19].

After labeling stage, different features are examined on determined regions. Therefore the essence of the labeled regions is specified. The main elements of graphical user interface have been considered in initial steps of this research. These elements are Checkbox, Textbox, Rectangular button, Circular button, Radio button. All these elements are well-formed and would not overlap.

For object recognition on Canvas, different features can be used such as: number of lines, number of equal lines, number of adjacent equal lines, number of 90-degree angles, number of detected labels on elements, compliance with square area, compliance with circle area, compliance with rectangular area, equality of distance between maximum and minimum points in direction X and direction Y of the region.

Two vertical and horizontal filters used for line detection are shown in (**Table 2**), (**Table 3**).

Table 2. Vertical filter for vertical line detection

| -1 | 2 | -1 |
|---|---|---|
| -1 | 2 | -1 |
| -1 | 2 | -1 |

Table 3. Horizontal filter for horizontal line detection

| -1 | -1 | -1 |
|---|---|---|
| 2 | 2 | 2 |
| -1 | -1 | -1 |

Recognized vertical and horizontal lines are stored in an array with start and end points coordinates. The difference of maximum width and height is the property which can distinguish circle and square from a rectangle [**19, 20**].

*Number of Labels on Designed Elements*

One of the features is the number of detected labels on the elements designed on Canvas. This feature allows better distinction between selected radio buttons and circular buttons, text box and rectangular button.

A combination of the labeling result and text drawing method of Canvas is used as a solution for this feature.

To make the sticking letters of words due to the used font type, intelligible using this mixed solution is highly favored.

Therefore in labeling phase, just one label is assigned to all letters. This leads to a higher probability of errors in element detection.

To obtain more precise results, the number of labels on the detected element is considered as a feature. Number of labels is considered as ten if the method detects just one label. If the number of detected labels is more than one, it will be equated with twenty, and if no label is detected, this amount would be equal to zero. Aim at this kind of selection is to standardize and increase capability of distinction in detecting elements.

*Region Area*

Region area is another feature, which helps detecting region essence. The value of this feature is calculated by counting region pixels once and again with the aid of region properties and formula area.

The results of two approaches are compared. The final result is useful in detecting region type.

*Number of 90 Degree Angle*

Another feature is number of 90-degree angle. Detected lines help to find the angles between two adjacent lines.

*Image Retrieval from Feature Base*

Content-based image retrieval technique describes visual content of images by multidimensional feature vector. These feature vectors form feature base. In the proposed extension, vector of sample images are stored in an array. As the results, the proposed extensions' feature base contains feature vectors of Textbox, selected check box, unselected check box, selected radio button, unselected radio button, rectangular button, circular button and feature vector of letters. The essence of input images is determined by calculating the similarities between the feature vectors of input images and those images of database. Described below is the formula for similarity measurement in Euclidean distance:

$$D(I, J) = (\sum_{c} |f_i(I) - f_i(J)|^p)^{\frac{1}{p}} \tag{1}$$

D (I, J), **Eq. (1)**, is as the distance measure between the query image I and the image J in the database, and $f_i(J)$ as the number of pixels in bin i of I. When p=1, 2 and ∞, D (I, J) is the $L_1$, $L_2$ (in this case L called Euclidean distance) and $L_\infty$ distance respectively [**18**].

*Supersede Canvas with Detected HTML Element*

The next phase supersedes Canvas with detected HTML elements. The extension replaces Canvas with a div element with the same width and height of Canvas. Furthermore, other detected elements would be replaced by equivalent HTML elements on div with the same coordinates. **(Fig 1)** displays a sample of equivalent HTML tag of detected designed element on Canvas.

```
<input role="checkbox" value="checkbox" name="checkbox1" tabindex="tabindex" onEvent="Event" id="elem1" type="checkbox" style="position: absolute ; left:leftpx ; top:toppx" />
```

Fig 1.Equal HTML element of designed unselected checkbox

WAI-ARIA techniques have been used in order to increase accessibility in superseded environment of Canvas.

The live-region has been added to this environment, which is one of the WAI-ARIA properties. Live-region property is used for dynamic content.

The superseded environment of Canvas operates through JavaScript API, thus the webpage would not be loaded again. As a result, the assistive technologies would not be aware of any changes on this superseded element. In addition, text on Canvas is not accessible, which should be changed into an accessible one. Text type in feature base has a feature vector.

Canvas has two methods for drawing on canvas, *fillText* and *strokeText*. Upon detection of text type from feature base, text coordinates' values, which had been detected from image processing phase, are compared with coordinates' arguments of drawing text method of Canvas.

If these two values were equal, the text would be equal with the first argument value of draw text method of Canvas.

Subsequently, text area is checked whether it is located in the area of another HTML element such as a button or not.

If the text was in the area of an element, it would be considered as the amount of element Value property. Otherwise it is added to the div as html label element. Each of the elements has a descriptive title, thus required information can be presented to the blind users more easily. Furthermore, the element tab index property is initialized considering their location and order of detection.

## 3.2 Mapping Canvas Methods and properties to Equal and common Javascript Codes:

For keeping Canvas functionalities in new superseded HTML environment, Canvas methods and properties must be mapped to equal and common JavaScript methods and functions.

Functionalities of Canvas methods or events can be divided into two forms. Either it operates regardless of mouse position on Canvas or different operations happened based on the mouse position on Canvas.

Detected Canvas method or event is assigned to newly produced HTML element event or method. Therefore when the function is a general operation regardless of mouse position, all elements would execute the operation. Otherwise with regard to the pointed coordinates, each element would operate differently. By substituting Canvas with div HTML element, Canvas methods and properties would be invalid. Thus code mapping has been used to solve this problem. Different Canvas methods are used for designing user interface elements. StrokeRect and fillRect are methods, used for designing intended elements such as checkbox, rectangular button and text box. Arc method is used for drawing circle so it can be employed for creating radio button or circular button. FillText and strokeText are used for creating text on Canvas element. Font and TextAlign properties are applied for adding style to canvas texts.

## 3.3 Keyboard Accessibility

Keyboard is the most-used tool for vision-impaired people to navigate through web environment. Moreover, one kind of functionalities of Canvas is executed considering mouse position.

Consequently, this project has proposed a solution to perform mouse functionalities through keyboard in new produced environment. For this purpose, *Keyup* event has been attached to div environment.

In this event, it is checked whether the pressed key is TAB key or not. Thus, the user can navigate by Tab key through different elements of div environment. Moreover, the coordinates of active element is caught and then the proper action takes place. All these functionalities and modifications happen without making any limitations on developers, and this is one of the other advantages of our extension.

### 3.4 Accessibility Rules Observance in Superseded Environment

Web accessibility rules are another important aspect of this extension, which are observed without any developers or users interference. Currently, WCAG 2.0 is the official standard of WAI and W3C, Which has 12 guidelines.

One of the instructions is to use elements, which are focusable by default, such as A, Area, Input and TextArea. The other advantage of using these elements is the accessibility of these elements by keyboard. Furthermore, screen readers interact with these elements more easily. The other instruction is about using labels for interactive elements, which provides descriptive information on these elements. The most important rule is keyboard accessibility, which the extension adds a method to new superseded environment for this purpose [**10, 21**].

## 4. Experimental Evolution

To examine the actual work of our implemented Firefox extension, two use cases which are investigated by the help of NVDA screen reader have been considered [**22**].

Both of them have been designed and tested for Navigable DOM proposal by W3C Canvas group. In order to clarify the problem space, it should be reminded that interactive designed graphical user interfaces on Canvas are the target for the proposed approach.

The first case contains two designed checkboxes on Canvas. Clicking on each one of them causes their correspondent html checkbox being clicked on web page.

By being completed loading page on Firefox browser and appearing Canvas, the extension detects Canvas. Image processing technique detects designed elements on Canvas and then replaces them with their html one. As shown in (**Fig**2), this use case provides the experimenting Shadow DOM or Navigable DOM proposal.

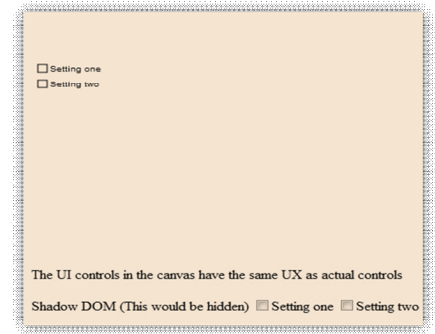

Fig2.The Case with designed check box on Canvas

Both of checkboxes keep their behavior after converting to their equivalent HTML element by extension.

Upon clicking on the converted HTML checkbox, the equivalent checkbox on the web page would be clicked. Moreover, this functionality is executable by keyboard, in other words; the new environment has keyboard accessibility (**Fig3**). In addition, the most important point is that in the newly produced HTML environment assistive technologies such as screen readers can interact much easier and transfer the semantic of this new section to blind people.

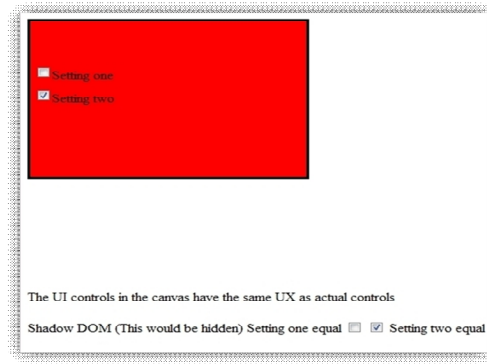

Fig3.Accessible superseded of Canvas element

Second use case contains three circular buttons, which by clicking each one of them, each button label changes (Fig**4**, Fig**5**).

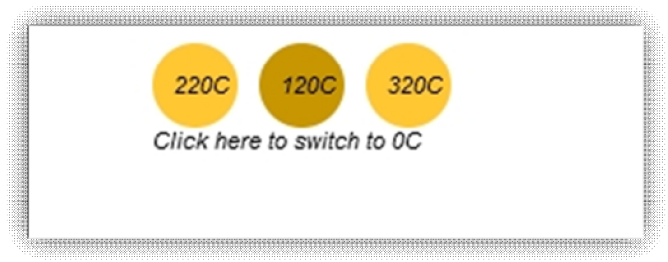

Fig4.The pattern used for experimenting extension

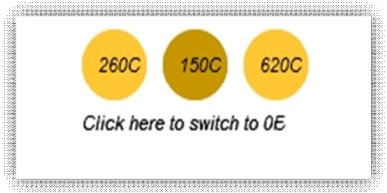

Fig5. Changes on user interface after clicking on Canvas

Demonstrated in (**Fig6**) is the result of converting Canvas interface by our Firefox extension.

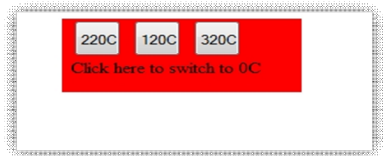

Fig6.Equivalent Accessible superseded environment of Canvas

By clicking on each button, labels' text changes, which stems from the designed Canvas functionality.

Moreover, by each change in labels, NVDA starts to read the labels from the change point, which is the result of using aria-live property. In addition, Keyboard accessibility rules are observed in this conversion.

## 4.1 Experimentation of Other Canvas Usage with Extension

Canvas has other usages besides designing a user interface, such as designing static images, games and dynamic graphics. In these cases the extension starts to process Canvas and try to detect elements as usual. If the extension found the expected elements, then the elements would be converted, otherwise the new produced environment would be just div element. (**Fig7**) is a sample of dynamic graphic on Canvas, which is converted to the (**Fig8**). As can be seen, little circles were detected as radio buttons.

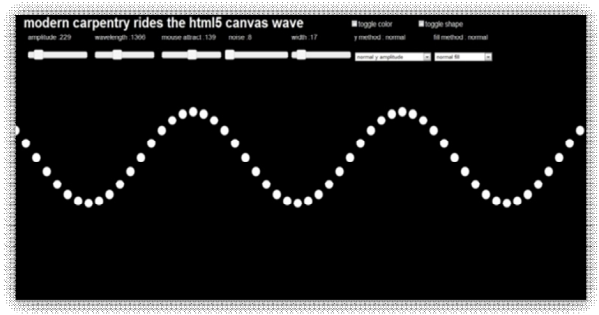

Fig7. Sample of designed dynamic graphic on Canvas

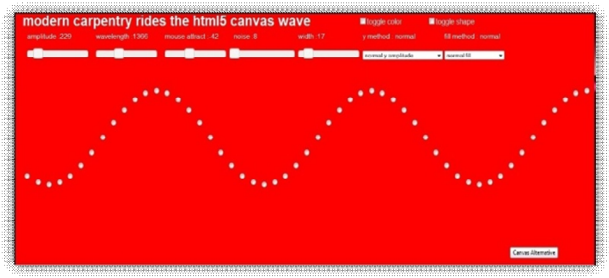

Fig8. Converted environment of Fig7 by extension

## 5. Conclusion and Future Work

Canvas is a new technology introduced in HTML5. It has different usages such as games, designing user interface, and dynamic graphics. Canvas inherently is inaccessible to assistive technology. For the purpose of making Canvas accessible, different groups are working. These groups are trying to modify Canvas HTML interface. Unfortunately, this modification imposes some limitations and rules on web developers and designers. Overlooking these rules can cause inaccessibility of Canvas.

In this paper, an extension designed on Firefox browser has been proposed. The proposal has different advantages. Firstly, it doesn't impose any rules and limitations on developers and designers. Secondly, there is no need of the user interference. Further, in the process of the proposed extension, accessibility rules are considered, especially keyboard accessibility has been added to supersede environment.

This extension detects Canvas in a web page followed by content-based image retrieval approach, which is an object detection technique capable of recognizing elements. In the next phase the extension converts detected element to equivalent HTML one.

Since Canvas replaced by new html environment, none of its methods and properties are valid. Therefore Canvas methods and properties are mapped to equivalent and common JavaScript instructions.

The most common elements in designing a user interface has been considered such as, radio button, check box, button, text box , circular button, with Texts on Canvas, which are inaccessible. The proposed extension converts texts on canvas to accessible text either as a label on div or value property of buttons. Keyboard accessibility and WAI-ARIA techniques are considered additionally. In the bigger picture, it seems that more elements can be added to this extension such as; combo box, links, progress and list box. With regards to not finalizing the development of the last version of HTML5 draft, more methods and properties would be added to Canvas. New methods and properties can improve extension capability and its functionality in code mapping and other interactions.

# 6. Conclusion and Future Work

Canvas is a new technology introduced in HTML5. It has different usages such as games, designing user interface, and dynamic graphics. Canvas inherently is inaccessible to assistive technology. For the purpose of making Canvas accessible, different groups are working. These groups are trying to modify Canvas HTML interface. Unfortunately, this modification imposes some limitations and rules on web developers and designers. Overlooking these rules can cause inaccessibility of Canvas.

In this paper, an extension designed on Firefox browser has been proposed. The proposal has different advantages. Firstly, it doesn't impose any rules and limitations on developers and designers. Secondly, there is no need of the user interference. Further, in the process of the proposed extension, accessibility rules are considered, especially keyboard accessibility has been added to supersede environment.

This extension detects Canvas in a web page followed by content-based image retrieval approach, which is an object detection technique capable of recognizing elements. In the next phase, the extension converts detected element to equivalent HTML one.

Since Canvas replaced by new html environment, none of its methods and properties are valid. Therefore Canvas methods and properties are mapped to equivalent and common JavaScript instructions.

The most common elements in designing a user interface has been considered such as, radio button, check box, button, text box , circular button, with Texts on Canvas, which are inaccessible. The proposed extension converts texts on canvas to accessible text either as a label on div or value property of buttons. Keyboard accessibility and

WAI-ARIA techniques are considered additionally. In the bigger picture, it seems that more elements can be added to this extension such as; combo box, links, progress and list box. With regards to not finalizing the development of the last version of HTML5 draft, more methods and properties would be added to Canvas. New methods and properties can improve extension capability and its functionality in code mapping and other interactions.

**Paniz Alipour Aghdam** received the B.Sc. in Computer Software Engineering from Azad university, Tehran Central Branch in 2007 and the M.Sc. from Azad university, Malayer Branch in 2012.Her main research interests are accessibility, web accessibility and HCI (Human Computer Interaction).

**Reza Ravanmehr** graduated as a computer engineer-hardware in 1996 from Shahid Beheshti University in Iran. After that he gained the master in 1999 from Azad University, Science and Research campus and also his PHD in computer-engineering in 2005. His main research interests are distributed/parallel systems and data mining. He is the faculty member of computer engineering department in Azad university, Tehran Central Branch from 2001.